\documentclass[11pt]{article}
\usepackage[dvips]{graphics}
\usepackage{times}

\def\titlepage{\@restonecolfalse\if@twocolumn\@restonecoltrue\onecolumn
 \else \newpage \fi \thispagestyle{empty}\c@page\z@}
\def\endtitlepage{\if@restonecol\twocolumn \else \newpage \fi}

\def\title#1{\begin{center}{{\bf #1}}\end{center}\smallskip}
\def\author#1#2{\begin{center}{\bf #1}\\ {\sl #2}\end{center}}
\def\abstract#1{\begin{quotation}\noindent{\small#1}\end{quotation}\medskip}
\def\caption#1{\noindent\small#1}
\def\email#1{\footnote{E-mail address: #1}}

\def\pacs#1{\vspace*{-3mm}\begin{quotation}\noindent{\small\sf PACS:} {\small #1}
\end{quotation}}
\def\@oddhead{\qquad{\shorttitle}\hfill\thepage}
\def\@evenhead{\thepage\qquad\hfill}
\def\thebibliography#1{
\baselineskip3mm
\itemsep0pt
\parsep10mm
\bigskip
\small
\centerline{\bf References}
\medskip
\list
 {[\arabic{enumi}]}{\settowidth\labelwidth{[#1]}\leftmargin\labelwidth
 \advance\leftmargin\labelsep\itemsep0pt
 \usecounter{enumi}}
 \def\newblock{\hskip .11em plus .33em minus .07em}
 \sloppy\clubpenalty4000\widowpenalty4000
 \sfcode`\.=1000\relax}

\begin{document}

\title{TWO-BODY RELAXATION TIMES IN HEATED NUCLEI}

\author{V.A. Plujko\email{plujko@mail.univ.kiev.ua}, O.M. Gorbachenko,
M.O.Kavatsyuk}
{Nuclear Physics Department,  Taras Shevchenko National University,
Pr. Acad. Glushkova, 2, bdg.11, 03022 Kiev, Ukraine}

\abstract{
The retardation and temperature effects in two-body collisions are studied.
The collision integral with retardation effects is obtained on the base of the
Kadanoff- Baym equations for Green functions in a form with allowance for
reaching the local equilibrium system. The collisional relaxation times of
collective vibrations are calculated  using both  the  transport approach and
doorway state mechanism with  hierarchy of particle-hole configurations in heated
nuclei. The  relaxation times of the kinetic method are rather slowly dependent
on multipolarity of the  Fermi surface distortion and  mode of the collective
motion. The dependence of the relaxation times  on temperature as well as on frequency
of collective vibrations is considered and compared. It is shown that
variations of the in-medium two-body cross-sections with energy lead to
non-quadratic dependence of the collisional relaxation time both on temperature
and on collective motion frequency.
}


\section{Introduction}

The damping of the collective excitations as well as transport
coefficients for viscosity and heat conductivity are strongly
governed by the particle  collisions. The relaxation time method
is widely used as the simplest and rather accurate approach for simulation
of the collisional relaxation rate $\lambda_{c}\propto 1/\tau$, where
$\tau$ is the so-called relaxation time
\cite{BS1970,kmpzp,AKoeh93}. Relaxation time method
can be  applied  to description of the decay rate of arbitrary mode of
motion, but an explicit form of the relaxation time depends on  specific
features of the mode. In this contribution, the collisional relaxation
times responsible for the width of the collective vibrations are studied.

\section{Semiclassical kinetic equation approach}

The collisional relaxation times can be calculated using the
collision integral of the transport equation. In studies of the damping
widths of collective excitation in the Fermi liquid, they are determined
by the coefficients $ \tau^{(\pm)}_{\ell}$ of the multipole expansion
of the total number ${\cal N}^{(\pm)}({\bf \hat{p}})$ of the
collisions in the direction ${\bf \hat{p}}={\bf p}/p$ of the momentum space
\cite{kmpzp,PLU99}
\begin{equation}
{\cal N}^{(\pm)}({\bf \hat{p}}) \equiv
\int^{\infty}_{0} d\epsilon J^{(\pm)}({\bf \hat{p}},\epsilon)
=\sum^{}_{\ell \ge \ell^{(\pm)}_{0}}
\sum^{\ell }_{m=-\ell }
{\phi^{(\pm)}_{\ell m}\over \tau^{(\pm)}_{\ell}}
Y_{\ell m}({\bf \hat{p}}) .
\label{number}
\end{equation}
Here, $J^{(\pm)}({\bf \hat{p}},\epsilon)$ are the linearized collision
integrals
\begin{equation}
J^{(\pm)}({\bf \hat{p}},\epsilon) \equiv J^{(\pm)}({\bf p},{\bf r}, t)  =
(J_{p}({\bf p},{\bf r}, t) \pm J_{n}({\bf p},{\bf r}, t))/2 ,
\label{numint}
\end{equation}
\noindent where the signs $(+)$ and $(-)$ stand for isoscalar and
isovector modes of vibrations, and the subscripts $p$ and $n$ stand for
protons and neutrons, respectively;  $J_{\alpha}({\bf p},{\bf r}, t)$ is a
collision integral in phase space $({\bf p},{\bf r})$, when a nucleon of the
sort $\alpha= ( p~{\rm or}~n )$ with momentum ${\bf p}$ is scattered;  $\epsilon$ is
nucleon energy. The collision integrals  are linearized with respect to the
dynamical component of the  phase space distribution function
$\delta f_{\alpha}({\bf p},{\bf r},t)$. The values $\ell^{(\pm)}_{0}$
determine the initial components of the multipole expansion of the total
number of the collisions. The functions $\nu^{(\pm)}_{\ell m}$  are the
partial components of the energy-integrated distribution function
$\delta f^{(\pm)}({\bf p},{\bf r},t) = ( \delta f_{p} \pm \delta f_{n})/2
\equiv \delta f^{(\pm)}({\bf \hat{p}},\epsilon,{\bf r},t)$,
\begin{equation}
\int^{\infty }_{0}d\epsilon \delta f^{(\pm)}({\bf \hat{p}},\epsilon,{\bf r},t)
= \sum^{}_{\ell \ge 0}\sum^{\ell }_{m=-\ell }
\phi^{(\pm)}_{\ell m}({\bf r}, t) Y_{\ell m}({\bf \hat{p}}) ,
\label{deltaf}
\end{equation}
where $Y_{\ell m}({\bf \hat{p}})$ is the spherical harmonic function.
In approximation of a small
difference in the chemical potentials for protons and neutrons
and assuming $\bar{f}_{p} = \bar{f}_{n} = \bar{f}$, where $\bar{f} \equiv
\bar{f} ({\bf p},{\bf r})$ is the equilibrium distribution function,
the dynamical distortions  $\delta f^{(\pm)}({\bf p},{\bf r},t)$
of the  phase space distribution functions are solutions of
the linearized  Landau-Vlasov equation
\begin{equation}
{\partial \delta f^{(\pm)}\over\partial t} +
{{\bf p }\over m} {\partial \delta f^{(\pm)} \over \partial {\bf r }}
- {\partial \bar{\epsilon}^{(\pm)} \over \partial {\bf r}}
{\partial \delta f^{(\pm)}
\over\partial {\bf p}} - {\partial \delta U^{(\pm)} \over \partial {\bf r}}
{\partial \bar{f} \over \partial {\bf p}} =
J^{(\pm)}({\bf p},{\bf r},t),
\label{h1}
\end{equation}
where $\delta U^{(\pm)} \equiv \delta U^{(\pm)} ({\bf p},{\bf r}, t)$ is
the Wigner transform of the variation of the self-consistent potential
with respect to the equilibrium value $\bar{\epsilon}^{(\pm)}$. In the nuclear
interior the mean field variation $\delta U^{(\pm)}$ can be expressed in
terms of the Landau interaction amplitude $F^{(\pm)}({\bf p}, {\bf p}\prime)$
as
\begin{equation}
\delta U^{(\pm)} = {g \over N_F} \int {d{\bf p}\prime
\over (2 \pi {\hbar})^3}\,\,
F^{(\pm)}({\bf p}, {\bf p}\prime) \,\,
\delta f^{(\pm)}( {\bf p}\prime,{\bf r}; t),
\label{h2}
\end{equation}
where $N_F = 2 \,p_F \,m^* /(g \,\pi^2 \,\hbar^3), \,\,\,p_F$ is the Fermi
momentum, $m^*$ is the effective mass of nucleon and $g$ is the spin
degeneracy factor.  The quantity $F^{(\pm)}({\bf p}, {\bf p}\prime)$
is usually parameterized in terms of the Landau constants
$F^{(\pm)}_0$ and $ F^{(\pm)}_1$ as
\begin{equation}
F^{(\pm)}({\bf p}, {\bf p}\prime) =  F^{(\pm)}_0 + F^{(\pm)}_1 
({\bf \hat{p}} \cdot {\bf \hat{p}}^\prime). 
\label{h3}
\end{equation}
In the isoscalar case, the Landau constants are related to the
incompressibility modulus $K$ \cite{blz} of matter and the
effective mass $m^*$ \cite{ring} by
\begin{equation}
K = 6 \mu (1 +  F^{(+)}_0), \ \ \ \ \ \ \ \
m^* = m \left(1 + F^{(+)}_1/3\right).
\label{h4}
\end{equation}
Here $m$ is the mass of free nucleon and $\mu$ is the chemical potential.
We have that  $\mu \approx\epsilon_F = p_F^2/2m^*$ for $T \ll \epsilon_F$,
where $\epsilon_F$ is the Fermi energy and $T$ is the temperature.
In the isovector case, the Landau parameter $F^{(-)}_0$ is related
to the nuclear symmetry energy $b_{\rm symm}$. Namely \cite{ktb,mlz},
\begin{equation}
b_{\rm symm} = {1\over 3}\,\mu \,(1 + F^{(-)}_0).
\label{h5}
\end{equation}

The quantities $\tau^{(\pm)}_{\ell }$ in Eq.(\ref{number})  can be
considered as the partial collective relaxation times because they determine
a collisional contribution to the damping widths resulting from the
two-body collisions in the layer of the momentum space with multipolarity
$\ell$,
\begin{equation}
{1\over \tau^{(\pm)}_{\ell }} \equiv  \int^{\infty }_{0}
d\epsilon \int d\Omega_p J^{(\pm)}({\bf \hat{p}},\epsilon)
Y_{\ell 0}({\bf \hat{p}}) \left/ \int^{\infty }_{0}d\epsilon \int d\Omega_{p}
\delta f^{(\pm)} Y_{\ell 0}({\bf \hat{p}}) \right. .
\label{taupar}
\end{equation}
These times  are
proportional to the relaxation times $\tau^{(\pm)}_{c}$ defining the
damping widths $\Gamma^{(\pm)}_{c}(L)$ of the isoscalar and the isovector
vibrations with frequency $\omega$ \cite{kmpzp,PLU99,KPS95,KPS96} in
regime of rare collisions with $\omega \tau^{(\pm)}_{c} \gg 1$ in the
Fermi liquid. In particular, the collisional damping widths of giant
resonances with dipole ($L=1$) and quadrupole ($L=2$) multipolarities
resemble the widths in the relaxation rate approach
\begin{equation}
\Gamma^{(\pm)}_{c}(L)=\hbar/\tau^{(\pm)}_{c}(L) , \ \ \
\tau^{(-)}_{c}(L=1) = \tau^{(-)}_{\ell=1}, \ \ \
\tau^{(+)}_{c}(L=2) = \tau^{(+)}_{\ell=2} ,
\label{width}
\end{equation}
in the case when nuclear fluid dynamical model with  relaxation is used
\cite{PLU99,KPS95,KPS96,plu2000a,plu2000b}. The collisional damping
width\cite{kmpzp} of zero sound in the Fermi liquid with its relative velocity
$S_{r}\simeq 1$ is also given by Eq.(\ref{width}) but with the use of the
$\tau^{(+)}_{\ell \to \infty} \propto \tau^{(+)}_{2}$ for
$\tau^{(\pm)}_{c}(L)$. The time $\tau^{(+)}_{\ell=2}$ at $\omega=0$
is the thermal relaxation time determining the viscosity
coefficient of the Fermi liquid~\cite{Bertsch78}.

The variations of the  mean field and of the  dynamical component of the
phase-space distribution function  change rapidly in the systems
with high frequency collective vibrations. This leads to the memory
(retardation, i.e. non-Markovian) effects in the collision term. There are
different expressions for memory-dependent collision integral  in the Fermi
liquid~(\cite{AS98}- \cite{mf99}).

The non-Markovian collision term of the semiclassical  Landau-Vlasov equation
was studied in  Born approximation with the use of the
Kadanoff- Baym equations for the Green functions in Refs.~\cite{kmp92,kp94}.
The one-component Fermi liquid was considered with the periodic time variation
of the nonequilibrium distribution function
$\delta f= \delta f_{n}= \delta f_{p}$,
$\delta f \propto \exp (-i\omega t)$.
As a result, the linearized collision integral consists of two
components (see Eqs.(42),(43) and (45),(46) of Ref.\cite{kp94}), i.e.,
\begin{equation}
J({\bf p}, {\bf r}, t) =
J^{(1)}({\bf p}, {\bf r}, t) + J^{(2)}({\bf p}, {\bf r}, t) ,
\label{eqJ}
\end{equation}
where the components $J^{(1)}({\bf p}, {\bf r}, t) $ and
$J^{(2)}({\bf p}, {\bf r}, t) $ are determined by the
variations of the distribution function and the mean field, respectively,
and
\begin{equation}
J^{(j)}({\bf p}, {\bf r}, t) = 2
\int {d{\bf p}_2\,d{\bf p}_3\,d{\bf p}_4\over (2\,\pi\,\hbar)^6}\,
W(\{{\bf p}_i\}) \delta (\Delta {\bf p}) \,B^{(j)} ({\bf p}, {\bf r}, t).
\label{eqsJj}
\end{equation}
Here, $W(\{{\bf p}_i\})=(d\sigma/d\Omega) 4(2\pi\hbar)^3/m^2 $ is the
probability  of two-body collisions with the initial momenta
${\bf p}_{1}= {\bf p}, {\bf p}_{2}$ and final ones
$ {\bf p}_{3}, {\bf p}_{4}$, $(i=1 \div 4)$;  $d\sigma/d\Omega$
is in-medium differential cross-section (in Born approximation);
$$
B^{(1)}({\bf p}, {\bf r}, t)=
\sum_{k=1}^4 \,\,\delta f_k(t) \,
{\partial \,Q(\{\bar{f}_j\})\over \partial \,\bar{f}_k} \,
[\delta_{+}(\Delta \bar{\epsilon} + \hbar \,\omega)
+ \delta_{-}(\Delta \bar{\epsilon} - \hbar \,\omega)],
$$
\begin{equation}
\label{eqsb12}
\end{equation}
$$
B^{(2)}({\bf p}, {\bf r}, t)=
Q(\{\bar{f}_j\}) {\Delta (\delta U(t)) \over \hbar\,\omega}
\{[\delta_{+}(\Delta \bar{\epsilon} + \hbar \omega)
- \delta_{+}(\Delta \bar{\epsilon})] -
[\delta_{-}(\Delta \bar{\epsilon} - \hbar \omega) -
\delta_{-}(\Delta \bar{\epsilon})]\},
$$
where $\bar{f}_{k} \equiv \bar{f}({\bf p}_{k},{\bf r} )$;
$\partial Q(\{\bar{f}_j\})/\partial \bar{f}_k$ are the derivatives of the
is the Pauli blocking factor $Q$ with respect to  $ \bar{f}_k$,
\begin{equation}
Q(\{\bar{f}_j\}) = (1-{\bar f}_{1})(1-{\bar f}_{2}){\bar f}_{3}{\bar f}_{4} -
{\bar f}_{1}{\bar f}_{2}(1-{\bar f}_{3})(1-{\bar f}_{4}) .
\label{Pauli}
\end{equation}
The ${\bar \epsilon}_{i}= \bar{\epsilon}({\bf p}_{i}, {\bf r})$ is the
equilibrium single-particle energy and  $\delta U_{j}$  the variation
of the mean field, and
$
\Delta \bar{\epsilon} = \bar{\epsilon}_{1} + \bar{\epsilon}_{2} -
\bar{\epsilon}_{3} - \bar{\epsilon}_{4},
$
$
\Delta (\delta U) \equiv \delta U_1 + \delta U_2 -
\delta U_3 - \delta U_4,
$
$ \Delta {\bf p}=
{\bf p}_{1}+ {\bf p}_{2}-{\bf p}_{3}-{\bf p}_{4}.
$
The  equilibrium distribution function $ \bar{f}_{k} \equiv
\bar{f} ({\bf p}_{k},{\bf r})$ depends on the equilibrium
single-particle energy  $\bar{\epsilon}_{k}
\equiv \bar{\epsilon}({\bf p}_{k},{\bf r})$:
$\bar{f}_{k} = {\bar f} (\bar{\epsilon}_{k})$. It equals the Fermi
function evaluated at the temperature $T$, $\bar{f}(\bar{\epsilon}_{k}) =
1/[ 1 + {\rm exp} ( ({ \bar{\epsilon}_{k} - \mu) / T })]$.
The nonequilibrium component $\delta f$ of the distribution function
can be presented in the
the form
\begin{equation}
\delta f({\bf p},{\bf r},t) = \, - \, \nu({\bf p},{\bf r},t)
{ \partial {\bar f}(\bar{\epsilon}) \over \partial \bar{\epsilon}}.
\label{df}
\end{equation}
With the use of this relation the expression for quantity $B^{(1)}$
can be transformed to  the form
\begin{eqnarray}
B^{(1)}({\bf p}, {\bf r}, t) &=& \, - \,
\sum_{k=1}^4 \,\,\nu_k \,
{\partial \,Q(\{\bar{f}_j\})\over \partial \,\bar{\epsilon}_k} \,
[\delta_{+}(\Delta \bar{\epsilon} + \hbar \,\omega)
+ \delta_{-}(\Delta \bar{\epsilon} - \hbar \,\omega)] =
\nonumber \\
&=&
\Delta \nu \,Q(\{\bar{f}_j\})
{\partial \,\over \partial \,\hbar \omega} \,
[\delta_{+}(\Delta \bar{\epsilon} + \hbar \,\omega)
+ \delta_{-}(\Delta \bar{\epsilon} - \hbar \,\omega)] - \delta B^{(1)} ,
\label{b1trans}
\end{eqnarray}
where $ \Delta \nu \equiv \nu_1 + \nu_2 -\nu_3 - \nu_4$,
$\nu_{k} = \nu ({\bf p}_{k},{\bf r},t)$ and
\begin{eqnarray}
\delta B^{(1)} &=&
\sum_{k=1}^4 \,
{\partial \over \partial \,\bar{\epsilon}_k} \,
\{ \nu_k Q(\{\bar{f}_j\}) [\delta_{+}
(\Delta \bar{\epsilon} + \hbar \,\omega)
+ \delta_{-}(\Delta \bar{\epsilon} - \hbar \,\omega)]\}
\nonumber \\
&+&
\sum_{k=1}^4 \,Q(\{\bar{f}_j\}) [\delta_{+}
(\Delta \bar{\epsilon} + \hbar \,\omega)
+ \delta_{-}(\Delta \bar{\epsilon} - \hbar \,\omega)]
{\partial \nu_k \over \partial \,\bar{\epsilon}_k} .
\label{b1corr}
\end{eqnarray}

The first component in the Eq.(\ref{b1corr}) determines a probability flux
of colliding particles which is connected with possibility of
variation of the energy $\bar{\epsilon}_{k}$ when the values of other
energies ($\bar{\epsilon}_{j \neq k}~{\rm and}~\hbar \omega$) are fixed. This term
should be equal zero because of fixing the total energy in initial or final
states and therefore it does not contribute to the total number of the
collisions ${\cal N}^{(\pm)}$ , Eq.(\ref{number}). The last statement
can be easily  verified by direct calculation of this contribution to
the ${\cal N}^{(\pm)}$ with the use of the procedure proposed by Abrikosov
and Khalatnikov (see Eqs.(\ref{angles})-(\ref{I34})) for evaluation of the
manifold energy integrals. A relative dynamical component $\nu_{k}$ of
the distribution function  is slowly dependent on energy and
it can be considered (at least for low temperatures $T \ll \epsilon_{F}$)
as a function of the momentum direction rather than
of the momentum: $\nu_{k} \equiv \nu ({\bf p}_{k},{\bf r},t) =
\nu ( {\bf \hat{p}}_{k},\epsilon_F, {\bf r},t)$.
Therefore the second component in the Eq.(\ref{b1corr}) is also negligible
and the term $\delta B^{(1)}$ in the Eq.(\ref{b1trans}) should be rejected,
\begin{equation}
\delta B^{(1)} = 0 .
\label{b1zero}
\end{equation}

Note that the generalized functions $\delta_{+}$, $\delta_{-}$ appearing in
Eqs. (\ref{eqsb12}), ({\ref{b1trans}) and ({\ref{b1corr}) include also
integral contribution,
\begin{equation}
\delta_{+}(x) = {1 \over 2 \pi}\,\int_{-\infty}^0 d\tau \,\,
e^{-i\,x\,\tau} = {i\over 2 \pi} \,{1\over {x + i0}} =
{1\over 2} \delta(x) - {1\over 2  \pi i} {\cal P}({1\over x}) , \ \ \
\delta_{-}(x)=\delta_{+}^{*}(x),
\label{genfun}
\end{equation}
where $\delta(x)$ is the delta function and the symbol ${\cal P}$ denotes
the principal value of integral contribution. The integral terms
of the  $\delta_{\pm}$, corresponding to virtual transitions, are usually
omitted in the $J$ because they assumed to be included by renormalizing
the interactions between particles \cite{ryt86}. This corresponds to
substituting $\delta (x)/2$ for $\delta_{\pm}$ in Eqs. (\ref{eqsb12}), i.e.,
to taking into account real transitions with conservation of energy.
The shift in energy $\Delta \bar{\epsilon}$ by  $\hbar \,\omega$ in the
arguments of the $\delta$-functions of the expressions for the collision
integral agrees with the interpretation of the collisions in the presence
of the collective excitations proposed by Landau \cite{LAND57}. According to
this interpretation, high-frequency oscillations in Fermi liquid can be
considered as phonons, that are absorbed and created at the two-particle
collisions.

In the one-component Fermi liquid the nonequilibrium distribution function
$\delta  f({\bf p},{\bf r},t)$ $ \equiv f({\bf p},{\bf r},t) -
{\bar f} (\bar{\epsilon}({\bf p},{\bf r}))$ is a solution of the
linearized Landau-Vlasov equation in the form
\begin{equation}
{ \partial \delta f \over \partial t}+ {p \over m}
( {\bf \hat{p}} \cdot {\partial \over \partial {\bf r}} )
\delta \bar{f} = J .
\label{LVone}
\end{equation}
Here, $\delta \bar{f}$ is a linear deviation of the distribution function
from its local equilibrium value $f_{l.e.}$, where a function $f_{l.e.}$
is equal to the  the Fermi function  ${\bar f} (\epsilon)$ evaluated
with actual one-particle energy $ \epsilon = {\bar \epsilon}+ \delta U$,
$f_{l.e.} = {\bar f} (\epsilon({\bf p},{\bf r},t))$,
\begin{equation}
\delta \bar{f}= \delta f - {d \bar{f} \over d \bar{\epsilon}} \delta U =
f({\bf p},{\bf r},t) -f_{l.e.} = \, - \,
\chi {d \bar{f} \over d \bar{\epsilon}}, \ \ \
\chi = \nu + \delta U ,
\label{flocal}
\end{equation}
where  relationship for $\delta U$ has the form of the Eq.(\ref{h2}) but
with the interactions and distribution function for one-component
Fermi liquid.

According to Eq.(\ref{LVone}), a Fermi system tends to the local equilibrium
(when $\partial \delta f / \partial t =0$) if the collision integral is a
functional of the $\delta\bar{f}$, $J =\Phi(\delta\bar{f})$. The collision
integral given by the expressions (\ref{eqJ})-(\ref{eqsb12}) and
(\ref{b1trans}), (\ref{b1zero}) has the following general form:
$\delta J=\delta J^{(1)}(\delta f)+\delta J^{(2)}(\delta U)$. Therefore
it does not lead to local equilibrium of a system. The condition of the
existence of the local equilibrium of a system is  general property of
the Landau- Vlasov equation in the Fermi liquid \cite{BP1991,LifPit} at
$\partial \delta f / \partial t = 0$.  Therefore the Born approximation
(\ref{eqJ})-(\ref{eqsb12}) for collision integral is poor approach
without additional modification, and a revision  of the
derivation method  of the  collision integral expression  is needed.

It should be initially noted that the foregoing relationships
for collision integral are obtained with the use of the perturbation
theory in  nearly nonhomogeneous systems with week interaction
between particles. A week interaction can not change  rapidly
the trajectory of the particle and due to this it  can not lead
to rapid variations of the  distribution function.
It means  that retardation effects are overestimated in the
expression for collision integral in Born approximation
where it was assumed that distribution function was varied very
quickly during all possible interval of the time changing
($ - \infty \le t^{\prime} \le t $). Consequently,
the collision integral given by Eqs. (\ref{eqJ})-(\ref{eqsb12}),
(\ref{b1trans}), (\ref{b1zero})  can be in fact correct in the case
of small retardation, i.e., for small values of the $\hbar \omega$.

With this in mind, we replace the derivatives of the form
$\partial \, \delta_{+}(\Delta \bar{\epsilon} + \hbar \,\omega) /
\partial \,\hbar \omega$ and
$\partial \, \delta_{-}(\Delta \bar{\epsilon} - \hbar \,\omega) /
\partial \,\hbar \omega$ in the Eq.(\ref{b1trans}) by the finite
differences $(\delta_{+}(\Delta \bar{\epsilon} + \hbar \,\omega) -
\delta_{+}(\Delta \bar{\epsilon})) /  \hbar \,\omega$ and
$(\delta_{-}(\Delta \bar{\epsilon}) -
\delta_{-}(\Delta \bar{\epsilon} - \hbar \,\omega))  /  \hbar \,\omega$,
respectively. Then we combine the resulting expression together with
contribution $B^{(2)}$ arising from mean-field variation  and obtain the
linearized collision integral for one-component Fermi liquid in the
following form
\begin{equation}
J({\bf p},{\bf r},t) =
 \int {d{\bf p}_2 d{\bf p}_3 d{\bf p}_4 \over (2\pi\hbar)^6}
  W(\{{\bf p}_i\}) \delta (\Delta {\bf p}) \Delta \chi Q
{\delta (\Delta \epsilon +\hbar\omega) - \delta ( \Delta \epsilon -\hbar\omega)
\over \hbar \omega } .
\label{int1}
\end{equation}
With the use of the algebraic relation \cite{BP1991}
\begin{equation}
[(1-{\bar f}_{1})(1-{\bar f}_{2}){\bar f}_{3}{\bar f}_{4}
- {\bar f}_{1}{\bar f}_{2}(1-{\bar f}_{3})(1-{\bar f}_{4})
\exp \left({\mp \hbar \omega \over T} \right)]
\delta (\Delta \epsilon \pm \hbar \omega)=0 ,
\label{ident}
\end{equation}
the Eq. (\ref{int1})  can be presented as\\

\noindent $ J({\bf p},{\bf r},t)= $
\begin{equation}
= \int{d{\bf p}_2 d{\bf p}_3 d{\bf p}_4 \over (2\pi\hbar)^6}
W(\{{\bf p}_i\})\delta(\Delta{\bf p})\Delta\chi
{\bar f}_{1}{\bar f}_{2}(1-{\bar f}_{3})(1-{\bar f}_{4})
\left[\Phi(\hbar\omega,T)-\Phi(-\hbar\omega,T)\right] ,
\label{int2}
\end{equation}
\noindent where $\Phi(\hbar \omega, T) = \delta(\Delta\epsilon+\hbar\omega)
[\exp(-\hbar \omega /T) -1]/ \hbar \omega$.

The collision integral  of the form (\ref{int1}) or (\ref{int2}) provides
driving distribution function towards its local equilibrium value because
it depends on the variation $\delta{\bar f}$, $J \equiv  J(\delta{\bar f})$.
This behaviour is in line with general properties of
the Landau- Vlasov equation in the Fermi-liquid \cite{BP1991,LifPit} at
$\partial \delta f / \partial t = 0$.
The expressions (\ref{int1}), (\ref{int2}) depend only on the occupation
probability  ${\cal P}_{2p2h}\equiv
{\bar f}_{1}{\bar f}_{2}(1-{\bar f}_{3})(1-{\bar f}_{4})$
of the 2p-2h states in the phase space. This fact leads to interpretation
of the collisional damping with linearized collision term as
the relaxation process due to the coupling of one-particle and one-hole
states to more complicated $2p-2h$ configurations.

The form of the collision integral (\ref{int2}) in the Markovian limit
($\omega \rightarrow 0$) coincides with the standard expression for
the collision integral in Fermi-liquid without retardation
effects~\cite{BP1991,LifPit} because in this case the term in square
brackets of Eq.(\ref{int2}) tends to the value
$- \delta(\Delta \epsilon) / T$.

The equation (\ref{int1}) for some special explicit form of the quantity
$\chi_{j}$ was used at first in Refs.~\cite{AB92,abb95,bab95}. The derivation
of the collision integral (\ref{int1}) is performed in Ref.~\cite{AS98} within
framework of the extended time-dependent Hartree-Fock model.
The expressions for the distortion functions $\chi_{j}$  corresponding to
a perturbation approach on collision term and including the
amplitudes of the random phase approximation were used in this method.

The expression for the collision integral in two-component Fermi-system
is obtained from Eq.(\ref{int2}) in the same manner as done in Ref.\cite{PLU99}
under the assumption that chemical potentials and  the equilibrium
distribution functions  are  the same for protons and neutrons.

The analytical expressions for partial collective relaxation times
$\tau_{\ell}$ ( Eq.(\ref{taupar} ) can be obtained in low-temperature and
low-frequency limits ($T, \hbar \omega \ll \epsilon_{F}$). In this case the
momentum integrals are calculated using the Abrikosov- Khalatnikov
procedure\cite{BP1991,AK58,SB1970,VVT85} which is based on the assumption that
particles are  scattered near Fermi surface with the momentum values $p_{i}$
approximately equal to the Fermi momentum  $p_{F}$. In this case
the probability $W(\{{\bf p}_i\})$ of two-body collisions  can be taken
as a function of two scattering angles $\phi$ and $\theta$, where
$\phi$ is the angle between the momenta ${\bf p}_{1}$
and ${\bf p}_{2}$, and  $\theta$ is the angle between the
$({\bf p}_{1} {\bf p}_{2})$ and  $({\bf p}_{3} {\bf p}_{4})$ planes;
that is
\begin{eqnarray}
\cos \phi &=& ({\bf \hat{p}}_2 \cdot {\bf \hat{p}}_1) ,
\nonumber \\
\cos \theta &=& [{\bf \hat{p}}_1\times{\bf \hat{p}}_2] \cdot
[{\bf \hat{p}}_3\times{\bf \hat{p}}_4]/
\vert[{\bf \hat{p}}_1\times{\bf \hat{p}}_2] \vert \vert
[{\bf \hat{p}}_3\times{\bf \hat{p}}_4]\vert .
\label{angles}
\end{eqnarray}
It allows to separate the angular and the energy integrations in the
collision integral at arbitrary scattering angle\cite{VVT85}.
The integrals with respect to momenta in expression for the collision
integral are calculated employing the transformation
\cite{BP1991,AK58,VVT85}
\begin{eqnarray}
&&\int d{\bf p}_{2}d{\bf p}_{3}d{\bf p}_{4}
\delta (\Delta {\bf p})(\ldots)=
\nonumber \\
&=&{m^{* 3}\over 2} \int^{\pi}_{0} d\phi
\int^{\pi}_{0}d\theta {\sin \phi \over \cos {\phi \over 2}}
\int^{2\pi}_{0}d\varphi \int^{\infty}_{0} \int^{\infty}_{0}
\int^{\infty}_{0} d\epsilon_2 d\epsilon_3 d\epsilon_4 (\ldots) .
\label{momtrans}
\end{eqnarray}
Here, $\varphi$ is the azimuthal angle of the momentum
${\bf p}_2$ in the coordinate system with the $Z$ axes along
${\bf p}_1$.

The integration with respect to the azimuthal angle $\varphi$
is performed by the relation\cite{SB1970}
\begin{equation}
\int^{2\pi}_{0}{d\varphi \over 2\pi} Y_{lm}({\bf \hat{p}}_{j})=
Y_{lm}({\bf \hat{p}}_{1})P_{l}({\bf \hat{p}}_{j} {\bf \hat{p}}_{1}) ,
\label{intphi}
\end{equation}
where $P_l$ is a Legendre polynomial, and
\begin{eqnarray}
({\bf \hat{p}}_{3} {\bf \hat{p}}_{1}) &=&
\cos^{2} (\phi/2) + \sin^{2} (\phi/2) \cos \theta ,
\nonumber \\
({\bf \hat{p}}_{4} {\bf \hat{p}}_{1}) &=&
\cos^{2} (\phi/2) - \sin^{2} (\phi/2) \cos \theta .
\label{addangles}
\end{eqnarray}
To perform over energies in the collision integral the following
expressions are used \cite{BP1991}
\begin{eqnarray}
I_{\nu}(y) &=& \int_{-\infty}^{+\infty} dx_1 dx_2 \ldots dx_{\nu}
n(x_1)n(x_2) \ldots n(x_{\nu}) \delta (x_1 + x_2 + \ldots + x_{\nu}) \equiv
\nonumber \\
&\equiv&\int_{-\infty}^{+\infty} dx_{\nu} n(x_{\nu}) I_{\nu-1}(x_{\nu}+y)
\equiv \int_{-\infty}^{+\infty} dt n(t-y) I_{\nu-1}(t) ,
\label{enerint}
\end{eqnarray}
where $n(x) = 1/( 1 + \exp {(x)} )$, $n(x) + n(-x) = 1$ and
\begin{equation}
I_3(y) = {1\over 2}{y^2 + \pi^2 \over 1 + \exp {(-y)}} , \ \ \
I_4(y) = {1\over 6}{y(y^2 + 4\pi^2) \over (1 - \exp {(-y)})} .
\label{I34}
\end{equation}

Finally we get the following relation for relaxation times using
the collisional integral of the form given by Eq.(\ref{int2}):
\begin{equation}
{\hbar \over \tau^{(\pm)}_{\ell}}=   {\cal R}(\omega, T)
\left[ <\sigma^{\prime}_{av} \Phi^{(+)}_{\ell }>+
< \sigma^{\prime}_{pn} \Phi^{(\pm)}_{\ell }> \right],
\label{atau}
\end{equation}
\noindent where $ \sigma^{\prime}_{av}=
( \sigma^{\prime}_{nn}+\sigma^{\prime}_{pp} )/2$;
$\sigma^{\prime}_{j j^{\prime}} \equiv d\sigma_{j j^{\prime}}/d\Omega$ is
in-medium differential cross-section for  scattering of the nucleons $j$ and
$j^{\prime}$ ( here, $j=n$ or $p$, and similarly $j^{\prime}= p$ or $n$).
The quantity ${\cal R}(\omega, T)$ result from the energy integrations and
has the following form in low-temperature and low-frequency limits 
($T, \hbar \omega \ll \epsilon_{F}$) in the approximation $m^{*} \simeq m$ 
\begin{equation}
{\cal R}(\omega, T)
=  {2\over 3 \pi} {m \over \hbar ^{2}}
\{ (\hbar \omega)^{2}+ (2\pi T)^{2}\} .
\label{rtau}
\end{equation}
The symbol $<\ldots >$ in Eq.(\ref{atau}) denotes averaging over angles
of the relative momenta of the colliding particles,
\begin{equation}
<(\ldots)> = {1\over \pi} \int^{\pi}_{0}
d \phi \sin(\phi /2) \int^{\pi}_{0} d \theta (\ldots) .
\label{angint}
\end{equation}

The functions $\Phi^{(\pm)}_{\ell}$ in (\ref{atau}) define the angular
constraints on nucleon scattering within the distorted layers of the
Fermi surface with multipolarity $\ell$:
\begin{equation}
\Phi^{(\pm)}_{\ell } = 1 \pm P_{\ell }(\cos {\phi}) -
P_{\ell }(({\bf \hat{p}}_{3} {\bf \hat{p}}_{1})) \mp
P_{\ell }(({\bf \hat{p}}_{4} {\bf \hat{p}}_{1})) ,
\label{phi}
\end{equation}
where the scalar products $({\bf \hat{p}}_{3} {\bf \hat{p}}_{1})$ and
$({\bf \hat{p}}_{4}{\bf \hat{p}}_{1})$ are given by Eq.(\ref{addangles}).
It follows
$
\Phi^{(+)}_{\ell=0}(\phi,\theta)=\Phi^{(+)}_{\ell=1}(\phi,\theta)=
\Phi^{(-)}_{\ell=0}(\phi,\theta) = 0 .
$
These relations lead to possibility of the two-body collisions in layers
of the Fermi surface distortion with multipolarity beginning with the value
$\ell^{(+)}_{0}=2$ in the isoscalar case and $\ell^{(-)}_{0} =1$ for the
isovector vibrations. As a result, the isovector
dipole relaxation time $\tau^{(-)}_{\ell=1}$ has  a finite
value, that means a nonconservation of the isovector current in the presence
of $n-p$ collisions~\cite{AIH83}.

Due to the momentum conservation and conditions $p_{i}$= $p_{F}$,
the angle $\theta$ agrees with the scattering angle in the center-of-mass
reference frame of two nucleons. The angle $\phi$ defines the magnitudes
of the relative momenta ${\bf k} _{i} = ({\bf p}_{2}-{\bf p}_{1})/2$
and  ${\bf k} _{f} = ({\bf p}_{4}-{\bf p}_{3})/2$ before and after collision,
respectively. The value of total momentum,
${\bf P}={\bf p}_{1}+{\bf p}_{2}$, also depends on a
magnitude of the $\phi$. We have
\begin{equation}
{\bf k}_{i}{\bf k}_{f} = k^{2} \cos \theta , \ \
k^{2} = k^{2}_{i} = k^{2}_{f} = p^{2}_{F} \sin^{2}(\phi/2) , \ \
{\bf P}^{2} = 4 p^{2}_{F} \cos^{2}(\phi/2) .
\label{reldef}
\end{equation}
Therefore the relative kinetic energy $E_{rel}$ of two nucleons as well as
the energy of centrum mass motion $E_{cm}$ are dependent on angle $\phi$
\begin{equation}
E_{rel} = k^{2}/m = 2 \epsilon_{F} \sin^{2}(\phi/2) , \ \ \
E_{cm} = P^{2}/2m = 2 \epsilon_{F} \cos^{2}(\phi/2)
\label{endef}
\end{equation}
and the total energy $E_{tot}=E_{rel}+E_{cm}$  holds only fixed,
$E_{tot}= 2 \epsilon_{F}$. Therefore the in-medium differential
cross-sections $\sigma^{\prime}_{j,m}$ of two nucleon scattering depend
on the relative momenta ${\bf k}_{i}$ and ${\bf k}_{f}$ at fixed total energy
rather then at fixed relative kinetic energy $E_{rel}$, because the magnitude
of $E_{rel}$ changes with angle $\phi$ between colliding particles.
The transfer momenta
${\bf q} = {\bf k}_{i} -{\bf k}_{f} = {\bf p}_{3} -{\bf p}_{1}$ and
${\bf q}^{\prime} = -( {\bf k}_{i} +{\bf k}_{f}) = {\bf p}_{1} -{\bf p}_{4}$
for scattering due to direct and exchange interactions respectively are
also functions of the $\phi$ and $\theta$:
$q = 2 k(\phi) \sin (\theta/2)$ and $q^{\prime} = 2 k(\phi) \cos (\theta/2)$.

Now we estimate the collisional relaxation times in the case of the
isotropic scattering with independent of energy the angle-integrated cross
sections $\sigma_{j j^{\prime}}$.
Performing angular integration in (\ref{atau}) with the use of
Eqs.(\ref{angint}) and (\ref{phi}) we find that
$1 / \tau^{(\pm)}_{\ell < \ell^{(\pm)}_{0}} = 0$   and
\begin{equation}
{\hbar \over \tau^{(\pm)}_{\ell}} =
{ 1 \over \alpha^{(\pm)}_{\ell} }
\left[(\hbar \omega/2 \pi)^2+T^2 \right] , \ \ \
{ 1 \over \alpha^{(\pm)}_{\ell} }={ 8 m\over 3 \hbar^{2}}
\left[ c_{\ell} \sigma_{av} + d^{(\pm)}_{\ell} \sigma_{np} \right] ,
\label{taucon}
\end{equation}
$$
c_{\ell}=1-{ 2-(-1)^{\ell} \over 2\ell+1} , \ \
d^{(-)}_{\ell}={1-(-1)^{\ell}\over 2\ell+1} , \ \
d^{(+)}_{\ell}=d^{(-)}_{\ell=0}=c_{\ell=0}=c_{\ell=1}=0 ,
$$
where $\sigma_{av}$ = $[\sigma_{pp} + \sigma_{nn} +2 \sigma_{np}]/4$ is
the in-medium spin-isospin averaged nucleon-nucleon cross section.
The magnitude of the in-medium  cross section $\sigma_{j j^{\prime}}$
is taken  usually proportional to the value of the
free space cross section $\sigma^{(free)}_{j j^{\prime}}$ with a factor
$F= \sigma_{j j^{\prime}} / \sigma^{(free)} _{j j^{\prime}}$, so that
the parameter $\alpha^{(\pm)}_{\ell}$ can be rewritten in the form
\begin{equation}
\alpha^{(\pm)}_{\ell}= \widetilde{\alpha}^{(\pm)}_{\ell} /F, \ \
\widetilde{\alpha}^{(\pm)}_{\ell}=
4.18 / \left[ c_{\ell} + 1.3 d^{(\pm)}_{\ell}\right], \ \ MeV.
\label{falpha}
\end{equation}
Here, the values $\sigma^{(free)}_{av} =3.75~fm^2$ and
$\sigma^{(free)}_{np} = 5~fm^2$ are adopted\cite{AB92,KPS96}; they
correspond to the free space cross sections near Fermi energy.

The relative relaxation times $\tau^{(\pm)}_{\ell}/\tau^{(+)}_{\ell=2}$
with the free space cross sections are shown  on Fig.1 in relation
to the multipolarity $\ell$ of the distorted layers of the Fermi
surface which are accessible to particle collisions. Solid and dashed
lines connect the values which correspond to isoscalar and isovector modes
of vibrations respectively.
The magnitudes of the relaxation times are different for isoscalar and
isovector modes of vibrations  and they are dependent on the multipolarity
$\ell$. As seen from the Fig.1, the collisional relaxation times  rather
slowly vary with multipolarity $\ell$ and with collective motion mode at
isotropic scattering with energy independent free cross sections.
In particular, parameters $\widetilde{\alpha}^{(\pm)}_{\ell}$, which
define relaxation times by the Eq.(\ref{taucon}),  take the same value
at ${\ell} \to \infty$, $\widetilde{\alpha}^{(\pm)}_{\ell=\to \infty} \equiv
\widetilde{\alpha} = 4.18~MeV$,
and
$\widetilde{\alpha} / \widetilde{\alpha}^{(-)}_{\ell=1} \simeq 0.9$,
$\widetilde{\alpha}^{(-)}_{\ell=2} /
\widetilde{\alpha}^{(-)}_{\ell=1} \simeq 1.1$;
$\widetilde{\alpha} / \widetilde{\alpha}^{(+)}_{\ell=2} \simeq 0.8$,
$\widetilde{\alpha}^{(+)}_{\ell=3} /
\widetilde{\alpha}^{(+)}_{\ell=2} \simeq 1.4$.
%
\smallskip
\\
\resizebox{\textwidth}{!}{\hbox{
  \resizebox{360bp}{252bp}{\includegraphics{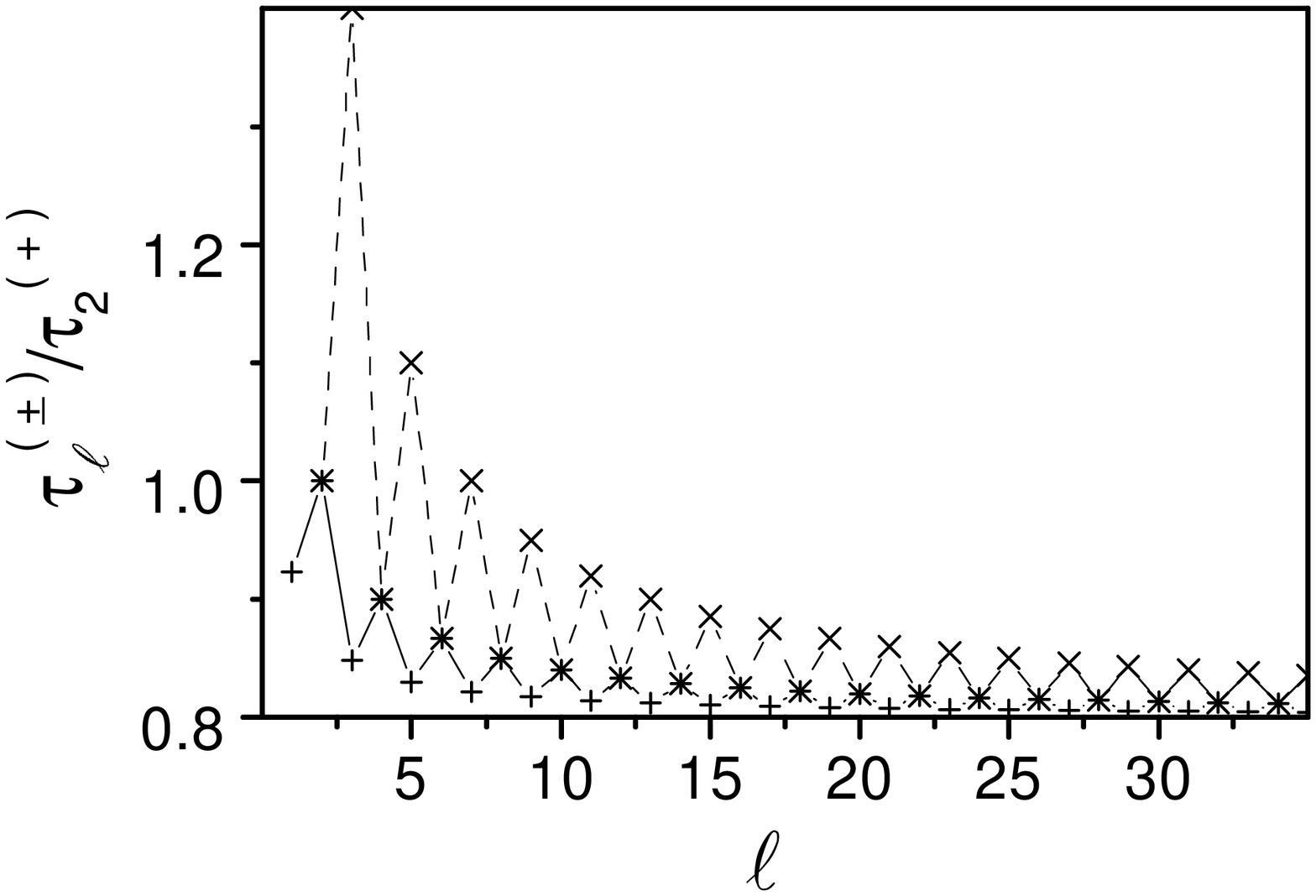}}
  \resizebox{360bp}{252bp}{\includegraphics{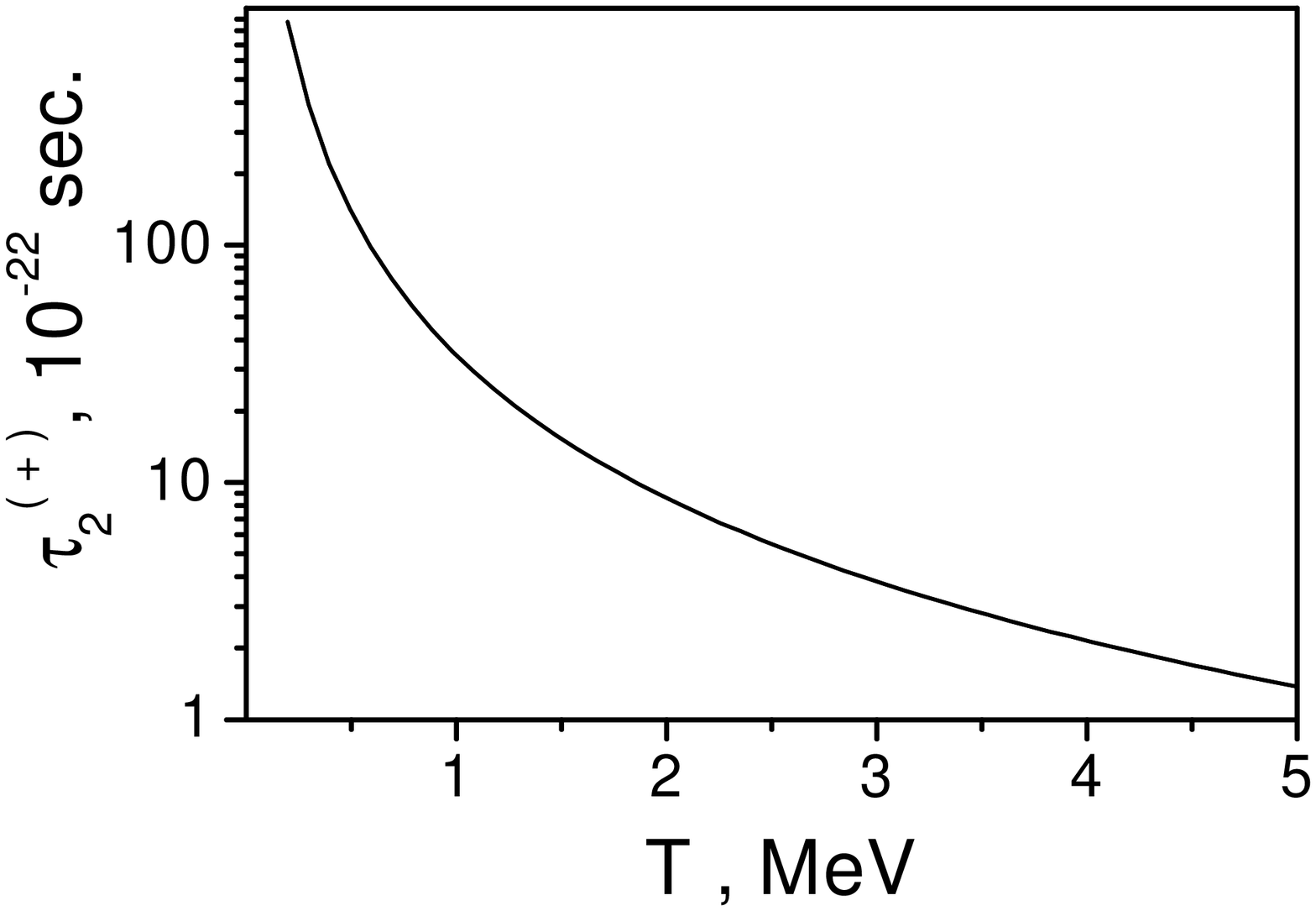}}
}}
\\\smallskip
\hbox to \textwidth{
\parbox[t]{0.48\textwidth}{Fig.1: The relative relaxation times $\tau^{(\pm)}_{\ell}
  /\tau^{(+)}_{2}$ versus multipolarity $\ell$.Solid and dashed
lines connect the values which correspond to isoscalar and isovector modes
of vibrations respectively.}
\hfill
\parbox[t]{0.48\textwidth}{Fig.2: The relaxation time $\tau^{(+)}_{2}$
in dependence of $T$ at $\hbar\omega=0$ in units of $10^{-22}$ sec.}
}

The dependence of the viscosity relaxation time $\tau^{(+)}_{\ell=2}(\omega=0, T)$
on the temperature is shown in Fig.2. The value of $\alpha^{(+)}_{\ell=2}$ is used
from Eq.(\ref{falpha}). The temperature dependence arises from smearing out the
equilibrium distribution function near the Fermi momentum in  heated nuclei.The
collisional rates $1 / \tau^{(\pm)}_{\ell}$ given by Eqs.(\ref{atau}) and (\ref{taucon})
are quadratic both in temperature and in frequency with the same relationship between
the components much as in the zero sound attenuation factor of heated Fermi liquid
within the Landau prescription \cite{LAND57,KPS95,AB92,AS98}.
The relaxation times $\tau^{(\pm)}_{\ell}$ depend on  frequency $\omega$ due to
the memory effects in the collision integral.

\section{Doorway state mechanism in heated nuclei}

The relationship (\ref{width}) gives the possibility to evaluate
the  relaxation time in system with weak damping in an independent way
from decay rates $\lambda^{(\pm)}_{c} = \Gamma^{(\pm)}_{c}/\hbar$.
We adopt the following physical notion:
$\lambda^{(\pm)}_{c} = \Gamma^{(\pm)}_{c}/\hbar$
is the spreading decay rate of the initial state
$\vert \nu^{(\pm)}_i \rangle$ to the final state
$\vert \nu^{(\pm)}_f \rangle$ within first- order approximation
of the time-dependent perturbation theory, as given by the golden rule
\begin{equation}
\lambda^{(\pm)}_{c} \equiv 1/ \tau^{(\pm)}_{c}
 ={2\pi \over \hbar}
\overline{\vert M^{(\pm)} \vert^2} \rho^{(\pm)}_{f},
\label{lambd}
\end{equation}
where $\rho^{(\pm)}_{f}$ is the density of the available final states.

The quantity $ \overline{\vert M^{(\pm)} \vert^2}$ is the mean
square matrix element for transitions  due to residual interaction $V_{res}$
\begin{equation}
\overline{\vert M^{(\pm)} \vert^2}=
\overline{\vert \langle\nu^{(\pm)}_f \vert V_{res} \vert \nu^{(\pm)}_i\rangle
\vert^2 } ,
\label{defm}
\end{equation}
where the line over symbols denotes an average over final
states \footnote{To simplify the presentation, we will omit in
the following the superscript $(\pm)$ and include them only when it is
necessary to avoid confusion.} \cite{DG65}.

The initial state should describe giant collective vibration in heated system
at given temperature $T$. It is taken as a mixture of a
collective state $(GR)$ and a thermal state which is approximated
by uncorrelated superposition states of $mp-mh$ configuration
with $m$ excited particles and holes corresponding to the most probable
number of excitons $\bar{n}=2m$ at given temperature $T$. The excitation
energy of the system is the sum of collective energy $\hbar \omega$ and
thermal excitation energy $U= \bar{n} \bar{\varepsilon}$ with
$\bar{\varepsilon}=\pi^{2}T/(12\ln2)$ for the average excitation energy per
thermal exciton~\cite{GH92}:
\begin{equation}
\vert \nu_i\rangle=\vert \{GR\}, \{mpmh\}\rangle, \
E=\hbar \omega+U,  \
U= aT^{2}, \
\bar{n}=2m=2 g T \ln2 ,
\label{inistate}
\end{equation}
where the expression for $\bar{n}$ is taken from \cite{GH92};
$a = \pi^2 g/6$. The quantity $g$ is the single nucleon state density
at the Fermi surface and the same values of $g$ are taken
for neutrons and protons.

Next we accept common feature that giant resonance  state $(GR)$
is formed by coherent superposition of many (predominantly correlated)
one-particle one-hole configurations and due to this fact wave function of
initial state can be presented as the sum of wave functions
$\vert  \{(m+1)p(m+1)h\}\rangle \equiv \vert  \{n_{i}\};k_{i}\rangle$
of  incoherent $(m+1)p-(m+1)h$ configurations  with $n_{i} = 2 + \bar{n}$
excitons, $k_{i}$ stands for other quantum numbers
\begin{equation}
\vert \nu_i\rangle = \sum_{k_{i}} C^{\nu_{i}}_{k_{i}} \vert
\{n_{i}\}; k_{i}\rangle  ,
\label{inisum}
\end{equation}
where the quantity $C^{\nu_{i}}_{k_{i}}$ defines the magnitude of the
admixture of different components of quasiparticle eigenstates.

Because of two-body character of the residual interaction $V_{res}$, the
final state can consist of  configurations with $n_{f}= n_{i}, n_{i} \pm 2$
excitons. The averaged squared matrix elements $\overline{\vert M \vert^2}$
of the transitions to states with fixed number of excitons can be rewritten
as
\begin{eqnarray}
&& \overline{\vert M \vert^2}=
\sum_{k_{i}, k^{\prime}_{i}} \overline{
C^{\nu_{i}}_{k_{i}} C^{\nu_{i},*}_{k^{\prime}_{i}}
\langle\{n_{f}\}; k_{f} \vert V_{res}\vert \{n_{i}\}; k_{i} \rangle
\langle\{n_{i}\}; k^{\prime}_{i} \vert V_{res}\vert \{n_{f}\}; k_{f} \rangle
} \simeq
\nonumber \\
&&\sum_{k_{i}}
\overline{ \vert C^{\nu_{i}}_{k_{i}} \vert^{2}
\vert \langle\{n_{f}\}; k_{f} \vert V_{res}\vert
\{n_{i}\}; k_{i} \rangle\vert^{2}} \simeq {\cal M}^{2}(n_{i} \to n_{f},E) ,
\label{msq}
\end{eqnarray}
where ${\cal M}^{2}(n_{i} \to n_{f},E) =
\overline{\vert \langle\{n_{f}\} \vert V_{res}\vert
\{n_{i}\} \rangle \vert^{2}}$
is  effective mean square matrix element for transition
between incoherent particle-hole states.

This transformation is performed by the use of the following
assumption and properties:\\
{i)} The compensation of the binary products of the matrix elements
coupling together incoherent exciton states with different values $k_{i}$ is
assumed to take place due to very complicated character of the final state;\\
{ii)} Approximate normalization of the factors  $C^{\nu_{i}}_{k_{i}}$
is used, $\sum_{k_{i}} \vert C^{\nu_{i}}_{k_{i}}\vert^{2} \simeq 1$;\\
{iii)} The  mean square matrix elements for transitions between
incoherent exciton states with different values of the numbers
$k_{i}, k_{f}$ are taken as equal to the same magnitude
${\cal M}^{2}(n_{i} \to n_{f},E)$ which is dependent only on numbers of
excitons and the total excitation energy. We also assume that
effective mean square matrix elements ${\cal M}^{2}(n_{i}\to n_{f},E)$
for interactions between different kinds of nucleons are equal in
magnitude~\cite{DB1983,K86,GH92,BH98}.

With the use of (\ref{msq}), the collisional relaxation rate
$\lambda_{c}$ (Eq.(\ref{lambd}))  coincides with the particle
interactions rate  of the exciton model starting from the $n_{i}$
configuration\cite{GH92}. The relation (\ref{lambd}) for the
collisional relaxation time $\tau_{c} = \tau_{c}( \omega, T)$ is
\begin{eqnarray}
&&{\hbar \over \tau_{c}} =
2\pi {\cal M}^{2}(n_{<}= \bar{n}+2 ,E)~\rho_{c}(E) +
2\pi {\cal M}^{2}(n_{<}= \bar{n} ,E)~\rho_{a}(E) ,
\label{tauexc}\\
&&E= \hbar \omega + U , \  \ U = a T^{2} , \ \ \
\bar{n} = b T , \ \  a=\pi^2 g/6 ,\ \  b= 0.843 a ,
\nonumber
\end{eqnarray}
when processes of  creation and of annihilation of the particle-hole
pairs are included. The matrix elements for both processes
are taken to be determined by the number excitons $n_{<}$ in the simplest
state\cite{K78} and $ \rho_{c}$ ($\rho_{a}$) is the density of the final
accessible states corresponding to the pair creation (annihilation).

The transitions to final configuration with $n_{f}= n_{i}+ 2 \equiv \bar{n}+4$
dominate at low excitation energies. Using
the simplest expression within the exciton model~\cite{GH92}
for density of final accessible states,
$\rho_{c}(E)= (g^3/ 2)(E^2 /(n_{i}+1))$, the Eq.(\ref{tauexc})
is given by
\begin{equation}
{ \hbar \over \tau_{c}( \omega, T)} =
\pi g^{3} {\cal M}^{2}(n_{<}= \bar{n}+2 ,E=\hbar \omega +U)
{(\hbar\omega+ a T^{2})^2 \over 3+ b T} .
\label{tauexcfin}
\end{equation}

According to the exciton model studies (\cite{K86}-\cite{Ob87}
the effective mean square matrix elements ${\cal M}^{2}(n_{i},E)$
is energy-independent at low excitation energies and it is
inversely proportional to energy at higher excitations.
The energy-independent estimation ${\cal M}^{2}$ was obtained with the
use of the Fermi gas model as\cite{BGMS72,GM73} ${\cal M}^{2} =
{\cal K}_{M} / A^{3}$, ${\cal K}_{M} \simeq 15.3~{\rm MeV}^{2}$, where $A$
is the mass number. The behaviour of collisional relaxation, as given by
Eq.(\ref{tauexcfin}), with such magnitude of the mean square matrix element
agrees  with estimation (\ref{taucon}) based on kinetic equation approach
at low temperatures $T \ll \hbar \omega$.

There are different estimates for the  mean square matrix element
with dependence on energy and number of excitons
\cite{K86,K78a,Ob87}. The fulfillment of the condition of
equiprobability of all particle-hole configurations is assumed in
most of them and therefore they can not be used in the considered
case of collective (predominantly 1p-1h) state overlapped with
temperature-fixed background particle-hole states. The expression
for ${\cal M}^{2}(n,E)$  without assumption on a uniform sharing
of the excitation energy $E$ into $n$ excitons was proposed in
Ref.~\cite{Ob87}:
\begin{equation}
{\cal M}^{2}(n_{<},E) = {n_{<}+1 \over 4}  {{\cal K}_{B} \over A^{3} E} ,
\label{pavel}
\end{equation}
where quantity ${\cal K}_{B}$ is not changed with $E$ and $n$ but can be
dependent on numbers of protons and neutrons,
${\cal K}_{B} = 190~{\rm MeV}^{3}$.  If this value of ${\cal M}^{2}$ is
employed as the squared intronuclear matrix element,  the collisional
relaxation time $\tau_{c}$ is a linear function of the collective energy
$\hbar \omega$ and thermal energy $U$. The corresponding expressions
(Eqs.(\ref{tauexcfin}) and (\ref{pavel})) for the relaxation time have the
same form as that one obtained within test particle approach, when collisions
were simulated by modeling s-wave scattering between
pseudoparticles~\cite{BBT89,SBT91}:
\begin{equation}
{ \hbar \over \tau_{c}( \omega,T)} =
{ 1 \over \alpha_{e} }
\left(\hbar \omega+ U \right) , \ \ \
{1 \over \alpha_{e}} = {{\cal K}_{B} \pi \over 4 } (g / A)^{3} .
\label{etauc}
\end{equation}
The relaxation times given by Eqs.(\ref{taucon}) and (\ref{etauc}) can
agree together at zero temperature if the magnitude of ${\cal K}_{B}$
is equal to the value ${\cal K}_{0}=(\hbar\omega / \alpha)(A/(g \pi))^{3}$.
Here, ${\cal K}_{0} \equiv 70.9 \hbar\omega / \alpha \simeq 220~{\rm MeV}^{3}$
for giant isovector dipole resonances  in heavy nuclei,
when $\hbar \omega \simeq 13~MeV$, $g=A/13$ and
$\alpha =\widetilde{\alpha} = 4.18~MeV$. This value of ${\cal K}_{B}$  is
rather close to the ${\cal K}_{B} = 190~{\rm MeV}^{3}$. It means that in
cold nuclei the relaxation times for the GDR within  doorway state mechanism
are not too different from those obtained  within  the  transport approach.

The dependence of the collisional relaxation times $\tau_{c}( \omega,T)$
on  temperature and energy $\hbar \omega$ is demonstrated  on Figs.3-6.
Solid and dashed lines correspond to the relaxation times
$\tau_{c}( \omega,T)$ within doorway state mechanism with the
mean square matrix elements ${\cal M}^{2}(n_{<},E) \propto 1/E $ and
${\cal M}^{2}(n_{<},E) = const $ respectively. Dot-dashed lines correspond
to the collisional relaxation times $\tau^{(-)}_{\ell=1}$ within  the
framework of the transport approach with the value $\alpha^{(-)}_{1}$
determined by  free $n-p$ cross-section ( see, Eq.(\ref{falpha})).
The factors ${\cal K}_{M}$ and ${\cal K}_{B}$ of the
mean square matrix elements are fixed from the condition of the
coincidence of the relaxation times $\tau_{c}( \omega = E_{GDR}/\hbar ,T=0)$ and
$\tau^{(-)}_{\ell=1}( \omega = E_{GDR}/\hbar ,T=0)$ in cold nuclei at a
frequency corresponding to the GDR energy. The magnitude of this energy
is taken as equals to the GDR energy in ${}^{208}Pb$: $E_{GDR} = 13.43~MeV$.
The values of the relaxation times are given in  units of $10^{-22}$ sec.

Figures 3, 4 show relaxation times at $\omega = 0$ (Fig.3) and
$\omega = E_{GDR}/\hbar$ (Fig.4) in relation to the temperature.
%
\\
\resizebox{\textwidth}{!}{\hbox{
  \resizebox{360bp}{252bp}{\includegraphics{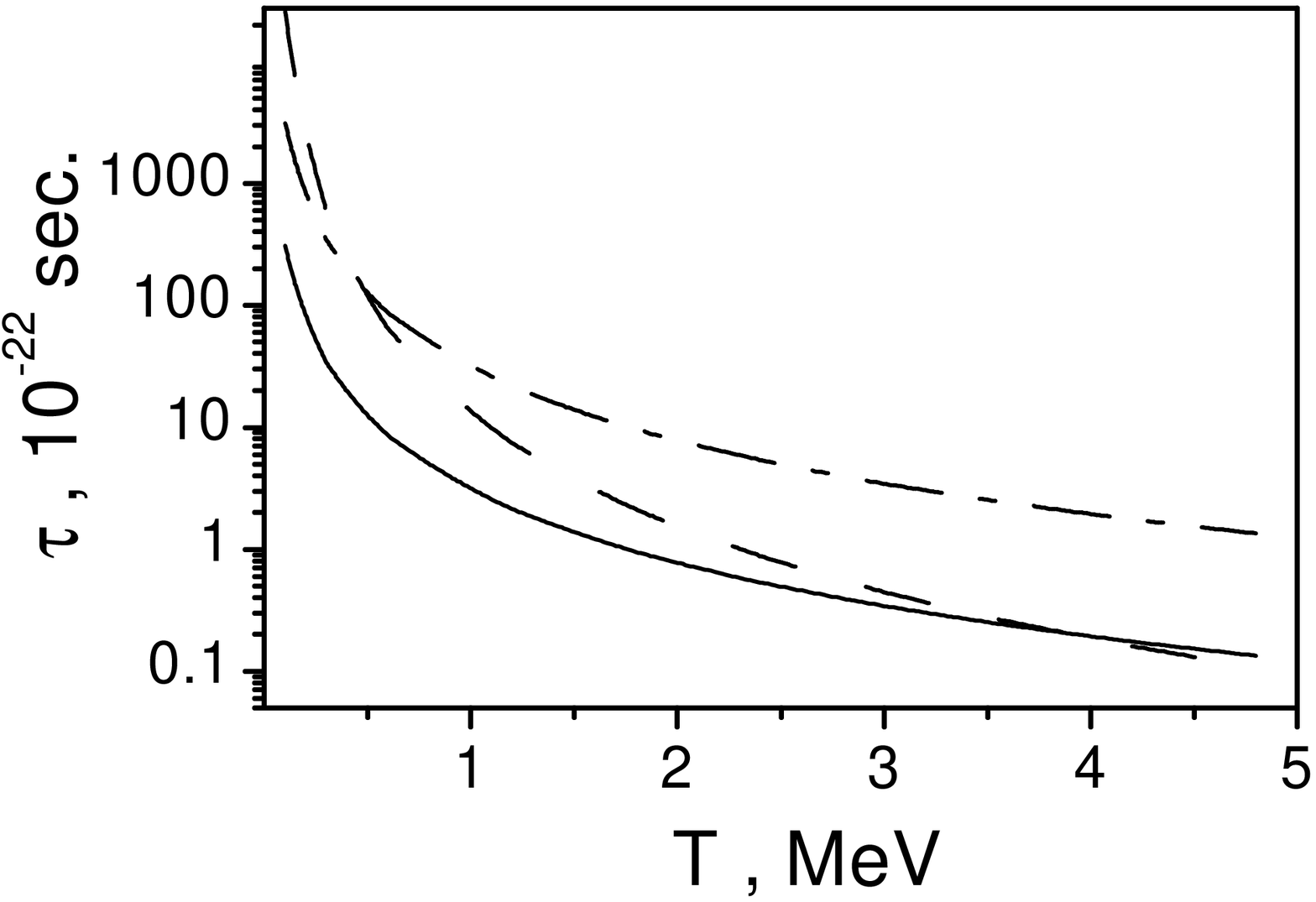}}
  \resizebox{360bp}{252bp}{\includegraphics{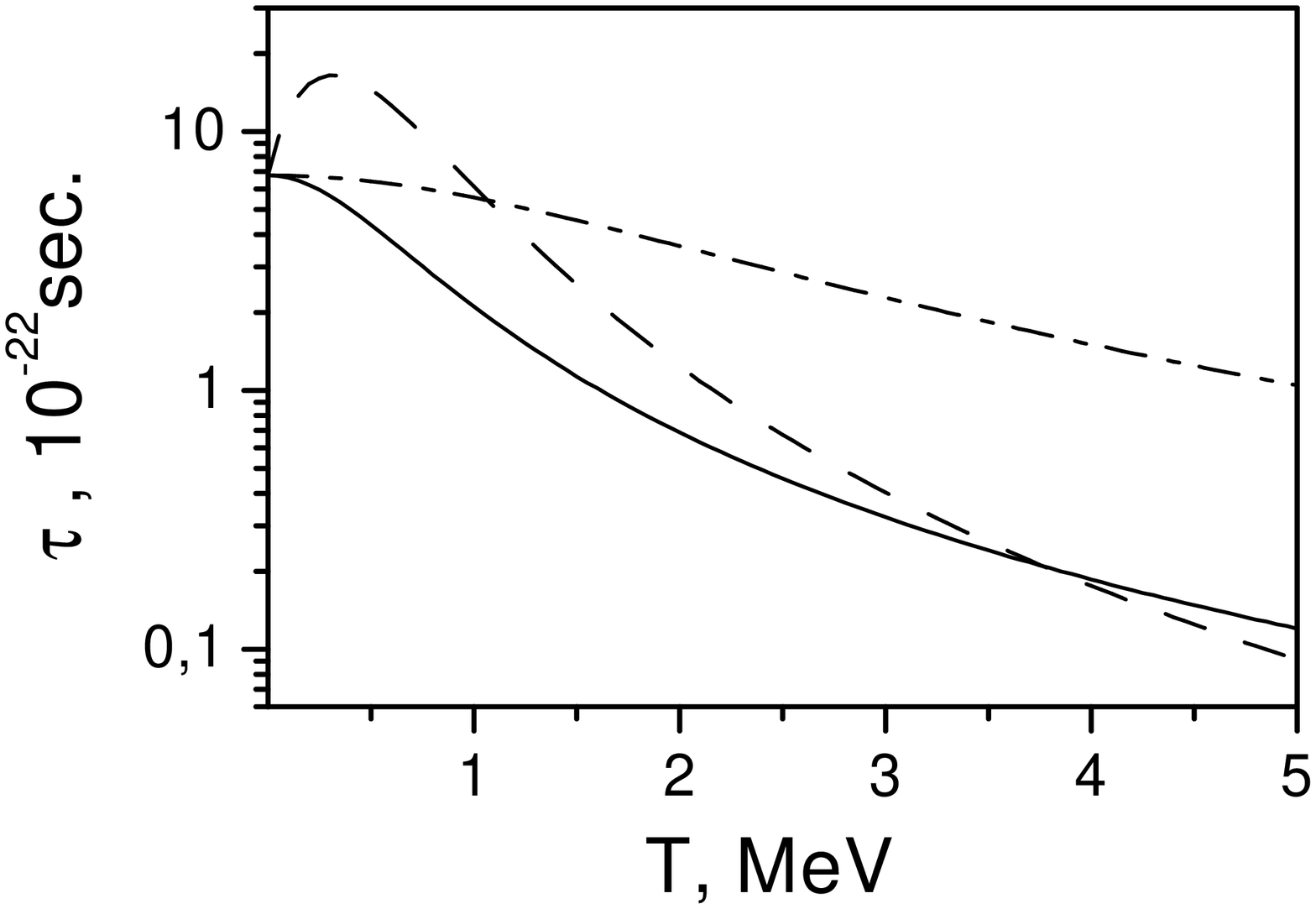}}
}}
\\\smallskip
\hbox to \textwidth{
\parbox[t]{0.48\textwidth}{Fig.3: The relaxation time $\tau$
in dependence of $T$ at $\hbar\omega=0$ in units of $10^{-22}$ sec.
Solid and dashed lines denote $\tau$ within dooway state mechanism with
$M^{2}\propto 1/E$ and $M^{2}\propto const$. Dot-dashed line is $\tau$ within
the framework of the transport approach.}
\hfill
\parbox[t]{0.48\textwidth}{Fig.4: The relaxation time in dependence of $T$ at
$\hbar\omega=E_{GDR}$ in units of $10^{-22}$ sec.
Notations are the same as in Fig.~3.}
}

Figures 5, 6 demonstrate dependence of the  relaxation times
on energy $\epsilon =\hbar \omega $  in cold (Fig.3, $T=0$) and
heated (Fig.4, $T=2~MeV$) nuclei.
%
\\
\resizebox{\textwidth}{!}{\hbox{
  \resizebox{360bp}{252bp}{\includegraphics{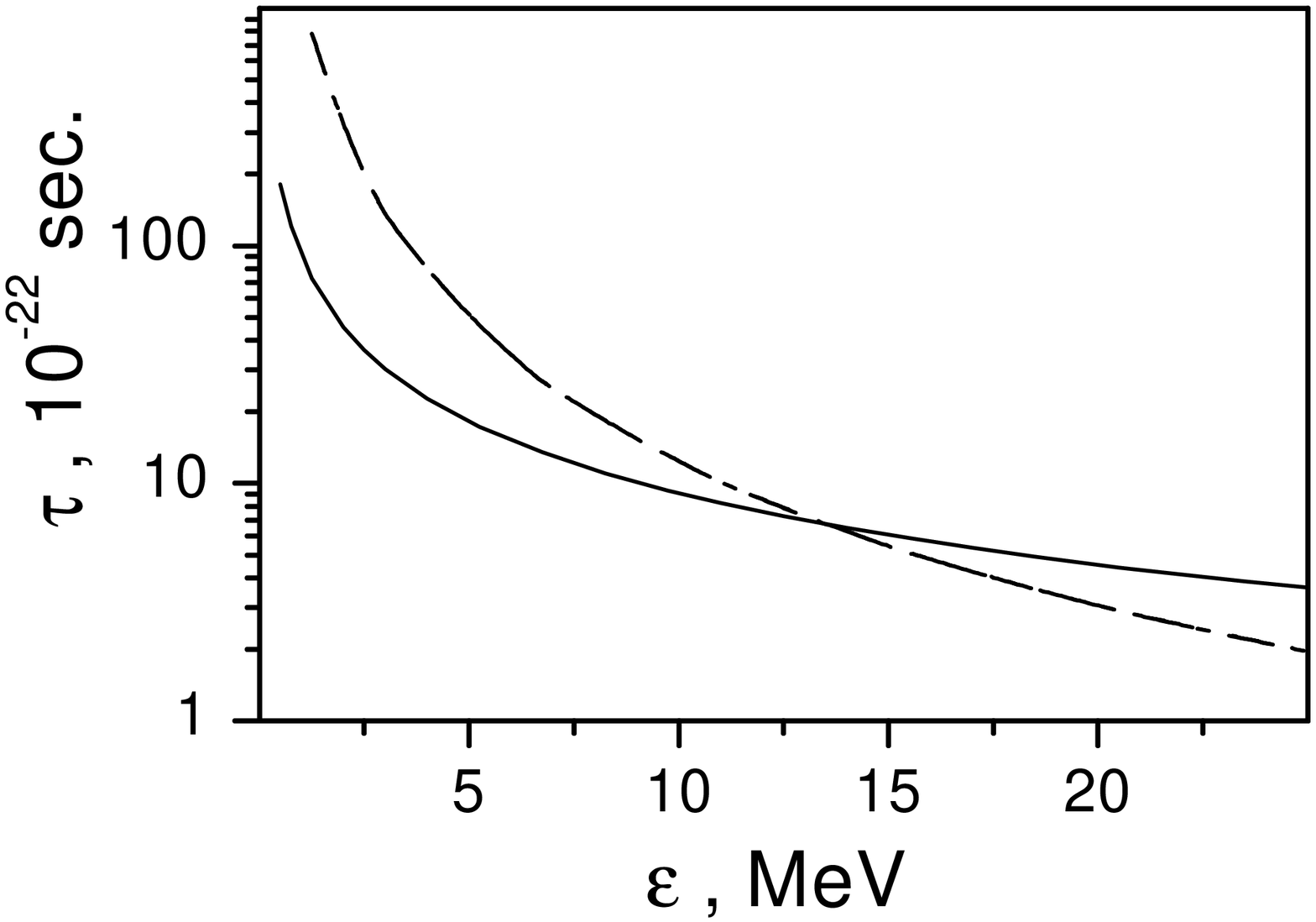}}
  \resizebox{360bp}{252bp}{\includegraphics{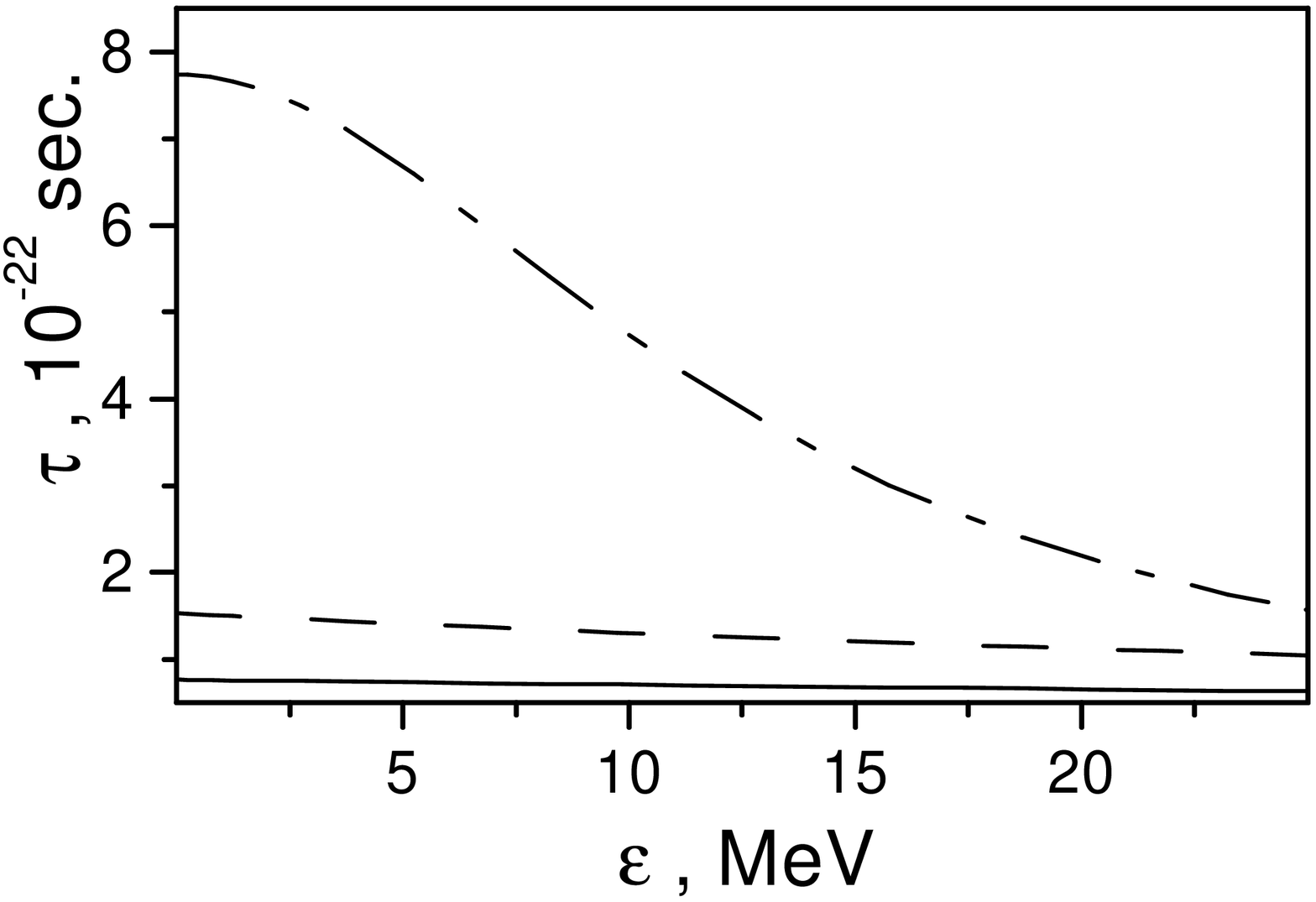}}
}}
\\\smallskip
\hbox to \textwidth{
\parbox[t]{0.48\textwidth}{Fig.5: The relaxation time
 in dependence of $\epsilon \equiv \hbar\omega$ at $T=0$ in units of $10^{-22}$ sec.
Notations are the same as in Fig.~3.}
\hfill
\parbox[t]{0.48\textwidth}{Fig.6: The relaxation time
in dependence of $\epsilon \equiv \hbar\omega$ at $T=2$ in units of $10^{-22}$ sec.
Notations are the same as in Fig.~3.}
}
\smallskip

The collective relaxation times in heated nuclei can be approximately
presented by the expression
\begin{equation}
1/\tau_{c}( \omega,T) =
q_1 [\omega^{\beta} + q_2  T^{2}]^{\gamma}/[q_3+q_4 T]^{\delta} ,
\label{apptau}
\end{equation}
where  $q_j$ are some constants and the exponents are the functions of
frequency and temperature and they are $\beta=2$, $\gamma=1$ , $\delta=0$
in the transport method; the $\beta$, $\gamma$ are changed from 2 to 1 and
$\delta$ varies from 1 to 0 with growing of the excitation energy in the
doorway state approach with allowance for pair creation.

\section{Results and conclusions}

The retardation and temperature effects in two-body collisions in heated Fermi-systems
were studied. An  expression for non-Markovian collision integral of the Landau-Vlasov
transport equation was obtained in a form which  allows for reaching the local equilibrium
in system. It was found in a small retardation limit on the base of the
Kadanoff- Baym equations for Green functions.

The  expressions for collisional relaxation times of  the collective vibration
in heated nuclei are derived with the use of the non-Markovian collision integral
as  well as  of the decay rates of exciton model. The relaxation times depend
on frequency of the collective vibrations and the temperature. The temperature
dependence arises from smearing out the equilibrium distribution function near
the Fermi momentum in  heated nuclei. The frequency dependence results from
the retardation (memory) effects in the  collisions. Analytical expressions for
relaxation times of the isoscalar and isovector modes of the collective motion
are derived in the case of the energy independent isotropic cross-sections in
the two-body collisions. The  relaxation times  rather slowly vary with multipolarity
of the Fermi surface distortions governed by collective motion and two-body
collisions. It gives possibility to use approximately the relaxation time ansatz
for collision integral. The relaxation times depends on type of collective motion
mode  like the lifetimes of the particle-hole configurations in two-component exciton
model of the Ref.\cite{DB1983}.

 New approach  for calculation of the collision relaxation time in heated nuclei
are proposed using the formulae for the transition rates of the
particle-particle transition between thermal state with collective vibrations and
incoherent particle-hole  configurations. This method leads to the same
results as the transport approach in the case of  low temperatures and energy
independent   mean square matrix element of interparticle collisions. It makes
possible to take into account the energy dependence of the in-medium cross-sections
in a simple phenomenological way by the use of the parametrization of the  mean
square matrix element ${\cal M}^{2}$ of interparticle collisions from  exciton model
of nuclear reactions.  The dependence of the matrix element ${\cal M}^{2}$ ( i.e.,
the in-medium cross-section) on energy  leads to non-quadratic dependence of the
relaxation times on temperature and collective vibration frequency.

\section{Acknowledgments}
This work is supported in part by the IAEA(Vienna) under contract
302-F4-UKR-11567.


\end{document}